\documentclass[lettersize,journal]{IEEEtran}
\usepackage{algorithmic}
\usepackage{array}
\usepackage[caption=false,font=normalsize,labelfont=sf,textfont=sf]{subfig}
\usepackage{textcomp}
\usepackage{stfloats}
\usepackage{url}
\usepackage{verbatim}
\usepackage{graphicx}

\usepackage{multirow,bbding}
\usepackage{amsmath,amssymb,amsfonts,bm}
\usepackage{threeparttable}

\usepackage{booktabs}

\usepackage[hidelinks]{hyperref}
\hypersetup{
  colorlinks=false
}

\usepackage{fontawesome5}

\usepackage{tikz}
\definecolor{lime}{HTML}{A6CE39}
\DeclareRobustCommand{\orcidicon}{%
    \begin{tikzpicture}
    \draw[lime, fill=lime] (0,0) 
    circle [radius=0.16] 
    node[white] {{\fontfamily{qag}\selectfont \tiny ID}};    \draw[white, fill=white] (-0.0625,0.095) 
    circle [radius=0.007];    \end{tikzpicture}
    \hspace{-2mm}}
\foreach \x in {A, ..., Z}{%
    \expandafter\xdef\csname orcid\x\endcsname{\noexpand\href{https://orcid.org/\csname orcidauthor\x\endcsname}{\noexpand\orcidicon}}
}

\def\BibTeX{{\rm B\kern-.05em{\sc i\kern-.025em b}\kern-.08em
    T\kern-.1667em\lower.7ex\hbox{E}\kern-.125emX}}
\usepackage{balance}
\begin{document}
\title{Learning Causality-inspired Representation Consistency for Video Anomaly Detection}

\author{
    \IEEEauthorblockN{
        Yang Liu\textsuperscript{1},
        Zhaoyang Xia\textsuperscript{2$^\dagger$}\thanks{$^\dagger$Equal contribution.},
        Mengyang Zhao\textsuperscript{1$^\dagger$},
        Donglai Wei\textsuperscript{1$^\dagger$},
        Yuzheng Wang\textsuperscript{1},
        Siao Liu\textsuperscript{1},
        Bobo Ju\textsuperscript{1},
        Gaoyun Fang\textsuperscript{1},
        Jing Liu\textsuperscript{1$^\ast$}\thanks{$^\ast$Corresponding authors.},
        Liang Song\textsuperscript{1$^\ast$}
    }

    \IEEEauthorblockA{
        \textsuperscript{1}Academy for Engineering \& Technology, Fudan University, Shanghai, China  \\
        \textsuperscript{2}Shanghai AI Laboratory, Shanghai China\\
        \{yang\_liu20, jingliu19, songl\}@fudan.edu.cn
    }
}

\maketitle

\begin{abstract}

  Video anomaly detection is an essential yet challenging task in the multimedia community, with promising applications in smart cities and secure communities. Existing methods attempt to learn abstract representations of regular events with statistical dependence to model the endogenous normality, which discriminates anomalies by measuring the deviations to the learned distribution. However, conventional representation learning is only a crude description of video normality and lacks an exploration of its underlying causality. The learned statistical dependence is unreliable for diverse regular events in the real world and may cause high false alarms due to overgeneralization. Inspired by causal representation learning, we think that there exists a causal variable capable of adequately representing the general patterns of regular events in which anomalies will present significant variations. Therefore, we design a causality-inspired representation consistency (CRC) framework to implicitly learn the unobservable causal variables of normality directly from available normal videos and detect abnormal events with the learned representation consistency. Extensive experiments show that the causality-inspired normality is robust to regular events with label-independent shifts, and the proposed CRC framework can quickly and accurately detect various complicated anomalies from real-world surveillance videos.

\end{abstract}


\section{Introduction}~\label{sec1}
Video anomaly detection (VAD) aims to automatically analyze the spatial-temporal patterns and contactlessly detect anomalous events of concern (e.g., traffic accidents, violent acts, and illegal operations) from surveillance videos \cite{liu2022icassp,liu2023amp}, which has promising applications in emerging areas such as traffic management \cite{yang2022novel,liu2022collaborative}, security protection \cite{wei2022msaf,liu2022abnormal}, and intelligent manufacturing \cite{song2022networking,ju2023high}. However, \textit{anomaly} is a vague concept, making anomalous events unbounded and difficult to predefine. Therefore, collecting all possible positive samples is impractical, making VAD remain challenging in the multimedia \cite{chen2022towards,liu2023generalized,yang2022emotion,yang2023novel,chen2022shape,yang2022learning,yang2023context,wang2022dpcnet,wang2022systematic} and pattern recognition \cite{yang2022disentangled,wei2022look,yang2023target,ju2023novel,chen2023query,liu2023improving,chen2023content,wang2023sampling,wang2022ferv39k} communities.

\begin{figure}
  \centering
  \includegraphics[width=\linewidth]{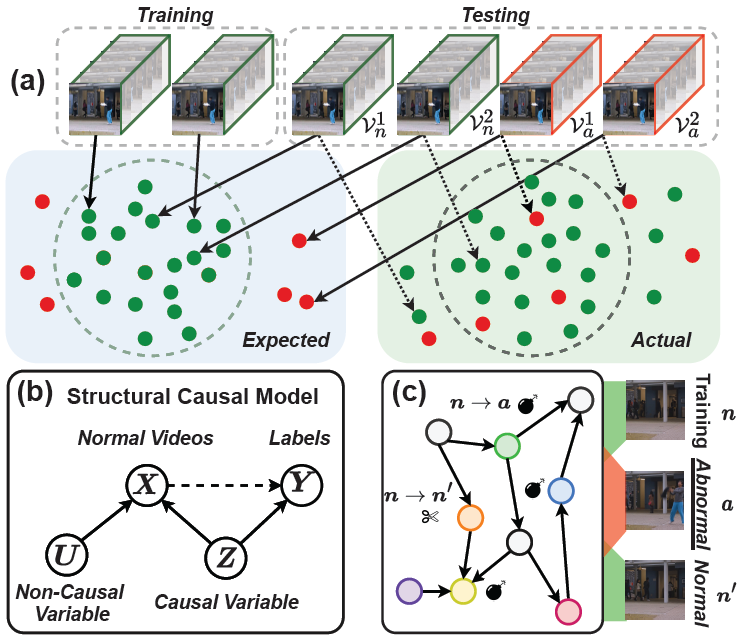}
  \caption{Motivation for addressing unsupervised VAD from a causality perspective. (a) illustrates the expectation (left) and the actuality (right) of the existing methods: they expect to train a characterizer to learn the pattern boundaries for regular events. However, the learned boundaries cannot effectively detect weak anomalies and unseen normal events. The structural causal model in (b) states the shortcomings of existing methods: they try to establish the statistical association (dashed arrow) of observable normal videos $\bm{X}$ with labels $\bm{Y}$, lacking effective exploration of causal factors. The sparse mechanism shift hypothesis in (c) suggests that label-independent domain shifts of diverse normal events ($\bm{n}\to \bm{n}^\prime$) have a limited and local impact on the learned causality (marked by {\footnotesize \faCut}). In contrast, anomalous events ($\bm{n}\to \bm{a}$) cause a full collapse of consistency (marked by {\footnotesize \faBomb}) learned on regular events. Inspired by the above observations, we expect to construct robust and efficient VAD models with CausalRL.}
  \label{motivation}
\end{figure}

To avoid the cost of collecting and labeling anomalous events, existing VAD methods typically use only normal videos to train a deep generative model (e.g., autoencoder \cite{hasan2016learning,liu2022appearance,wang2023adversarial}, generative adversarial network \cite{liu2022learning}, and transformer \cite{sun2023transformer}) to perform reconstruction or prediction tasks, which learns the distribution dependence of regular events in an unsupervised manner and treating out-of-distribution samples as anomalies. They assume that models learned on negative samples cannot characterize unseen positive ones, leading to significant deviations from the learned normality. As shown in Figure~\ref{motivation}(a), these methods \cite{liu2021learning,liu2023msn,cheng2023spatial} expect to map regular events (green cubes) to a hyperspace, while uncharacterizable anomalous events (red cubes) will fall outside. Benefiting from deep representation learning (DeepRL) \cite{liu2023dsdcla,xiang2023two,liu2023distributional,yang2023aide,yang2023spatio}, unsupervised VAD has achieved remarkable progress in recent years.

However, the long-overlooked problem is that regular events are also diverse. Besides, due to the high dimensionality of real-world videos and the complexity of target-scene interactions, regular events contain both shared and private semantics, i.e., prototypical and personalized features \cite{wu2022self}. In addition, positive and negative samples captured from the same scene usually share the most appearance context, making it difficult to filter these task-irrelevant semantics under an unsupervised setting. Ultimately, an unaffordable consequence is that the learned model may represent abnormal events well due to the overgeneralization of deep neural networks \cite{gong2019memorizing}, leading to missed detections, as sample $\mathcal{V}_a^1$ in Figure~\ref{motivation}(a). Moreover, for regular events with label-independent distribution offsets (e.g., the color of pedestrians' clothes and their walking posture in crowd anomaly detection), existing unsupervised methods cannot resist such random disturbances due to insufficient robustness, resulting in high false alarms, as sample $\mathcal{V}_n^1$ in Figure~\ref{motivation}(a). Therefore, existing methods are limited by DeepRL and only establish the crude statistical dependence of the normal distribution, making the learned normality unable to cope with complicated anomalies and normal events with unseen bias.

Inspired by causal representation learning (CausalRL) \cite{scholkopf2021toward}, we attempt to learn task-specific representations that contain potential causal mechanisms capable of revealing the intrinsic properties of regular events, which can mitigate the negative impact of event diversity and random label-independent bias in unsupervised normality learning \cite{li2022adaptive}. In this regard, we construct a structural causal model (SCM) shown in Figure~\ref{motivation}(b), where $\bm{X}$ and $\bm{Y}$ denote observable normal videos and labels, respectively. According to the common cause principle, we consider that there are causal factors in $\bm{X}$ that can fully describe normality. We attempt to learn these unobservable causal factors with the label consistency between shared and private features of $\bm{X}$. In addition, the sparse mechanism shift hypothesis point out that diverse normal events with significant distribution differences only vary locally in the high-level causality space. This hypothesis suggests that learning causality-inspired normality may enable the model to correctly infer shifted regular events and reduce false alarm rates. To this end, we propose an end-to-end causality-inspired representation consistency (CRC) framework to mining causal variable for unsupervised video anomaly detection. In the training phase, we optimize the CRC framework using the causal independence and the consistency of the multi-view representations for regular events. While testing, the causality-inspired characterizer learned on negative samples will not work for anomalies, making the anomalous events show significant differences in causal consistency, as shown in Figure~\ref{motivation}(c).

Specifically, the CRC framework first utilizes a DeepRL-based feature extractor to obtain original spatial-temporal patterns containing causal and non-causal variables. Then, an iteratively updated memory pool \cite{park2020learning} is used to record the general pattern of regular events. Distinguishing from existing methods that use features retrieved from the memory pool as representations for anomaly discrimination, we introduce a prototype decomposer to split the shared and private features \cite{wu2022self}. The shared and private features come from the same batch of training samples, so both are representations of regular events by nature. Therefore, we use causal factorization to reduce them into a set of causal factors and capture the intrinsic causal mechanism. Finally, the features represented by causal factors are fed into a clustering algorithm \cite{chang2020clustering} to obtain compact task-specific causal representations. The main contributions of this paper are summarized as follows:
\begin{itemize}
  \item We address the unsupervised video anomaly detection from a causality perspective and propose a causality-inspired representation consistency framework to learn video normality and detect anomalous events by consistency.
  \item We design prototype decomposer and causal factorization to mine causal variable directly from the videos and use causal representations to characterize normal events.
  \item Our method can cope with label-independent shifts and amplify the deviation of subtle anomalies with causal consistency. Experimental results prove the effectiveness of CRC, which achieves superior performance on benchmarks.
\end{itemize}

\section{RELATED WORK}
\subsection{Video Anomaly Detection}
VAD has been extensively studied for years due to its potential applications in emerging fields such as smart cities and secure communities, and various routes have been derived with the development of DeepRL. Among them, unsupervised methods \cite{hasan2016learning,cai2021appearance} follow the open-world assumptions without predefining and collecting anomalies and avoiding annotation costs and data imbalance problems, becoming the preferred solutions. Early unsupervised methods \cite{kim2009observe,ped2} treat VAD as a one-class classification task and use OC-SVM, OC-NN, or deep clustering to determine the boundaries of manual features, which are prone to the curse of dimensionality when dealing with real videos. Benefiting from the rise of deep generative models, researchers used autoencoders \cite{chang2020clustering} and generative adversarial networks \cite{liu2018future,liu2022learning} to extract spatial-temporal representations and introduce proxy tasks to learn the prototypical pattern of normal videos, i.e., normality. Specifically, such methods assume that generative models trained with normal videos are only effective in representing regular events. Thus, positive samples will experience significant performance degradation on the proxy task during the downstream anomaly detection phase. For example, Hasan \textit{et al.} \cite{hasan2016learning} propose a 2D convolutional autoencoder to reconstruct input sequences and use the reconstruction error to compute anomaly scores. Liu \textit{et al.} \cite{liu2018future} pioneered a video prediction framework to learn video normality and measure the degree of anomalies with the appearance and motion prediction errors. Following efforts lie in structure modifications (e.g., using dual-stream networks to learn appearance and motion normality separately \cite{chang2020clustering,chang2022video}) and proxy task stacking \cite{nguyen2019anomaly}. 

Recently, the intrinsic semantics consistency between different dimensions \cite{cai2021appearance} or regions \cite{liu2023osin} is considered feasible for video normality learning. Drawing that memory networks \cite{gong2019memorizing,park2020learning} can store prototype patterns of training samples, Cai \textit{et al.} \cite{cai2021appearance} construct two memory-enhanced autoencoders to learn appearance-motion consistency by learning relationships between RGB images and optical flow. They argue that the consistency learned on normal samples holds only for regular events. In addition, emerging object-level schemes \cite{chen2022comprehensive,liu2023osin,bao2022hierarchical} attempt to explore normal target-scene semantics interactions and discriminate anomalies accordingly. However, due to the high dimensionality and complexity of videos, both normal and abnormal events are diverse. Conventional DeepRL struggles to obtain representations with sufficient discrimination to describe diverse regular events. Studies show that such methods may miss-detect positive samples due to overgeneralization or fail to effectively reason about unseen negative samples due to insufficient representation ability. In contrast, our proposed CRC framework learns video normality with CausalRL, aiming to balance representation and generalization with causal mechanisms. Although causality has been proven practicable in numerous image processing tasks \cite{lv2022causality}, the one-class classification setting that only negative samples are available for training in VAD makes designing reasonable causal interventions and factorization schemes for casual variable extremely challenging.

\subsection{Causal Representation Learning}
Conventional deep representation learning has flourished in recent years with the emergence of large-scale multi-source datasets. However, DeepRL only learns the statistical independence between training samples and given labels, subject to the independent and identically distributed (i.i.d) assumption \cite{scholkopf2021toward}. Causal representation learning considers statistical independence as a crude description of the physical world, which cannot perform correct inference under distribution changes and intervention conditions. Currently, causality-driven representation models achieve leading performance in various applications such as domain generalization \cite{lv2022causality} and online recommendation systems, demonstrating great potential for learning robust representations and reusable mechanisms, which are also the key to high-performance VAED. On the one hand, real-world normal videos suffer from unpredictable bias, i.e., regular events contain personalized semantics that does not constitute anomalies. DeepRL-based characterizer can hardly accommodate such private features and may misclassify unseen regular events as anomalies. On the other hand, the unbounded nature of anomalous events makes it inevitable that their patterns intersect with the normal distribution. To summarize, robust anomaly detectors need to learn representations containing essential factors that adequately describe normality, which motivates us to learn normality with CausalRL.

\begin{figure*}
  \centering
  \includegraphics[width=\textwidth]{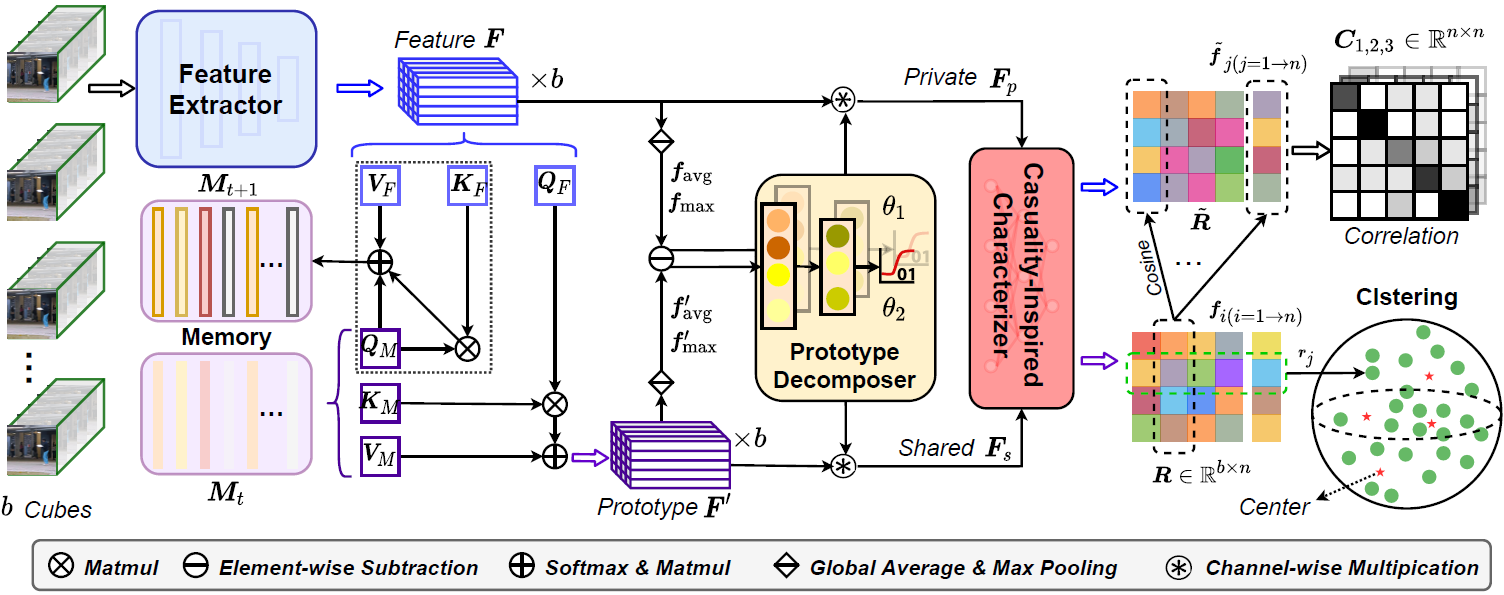}
  \caption{Overview of the proposed causally-inspired representation consistency (CRC) framework. In the training phase, CRC extracts the spatial-temporal features $\bm{F}$ of $b$ normal sequences and stores the prototypes in the memory pool $\mathcal{M}$, and then uses the prototype decomposer to strip the private and shared features $\{\bm{F}_p, \bm{F}_s\}$, which are fed to the causally-inspired characterizer (CiC) to learn the causal variables. Inspired by \hyperref[s1]{\S1} and \hyperref[s2]{\S2}, we exploit the independence of the causal variables to compute the correlation matrices $\bm{C}_{1,2,3}$ and optimize CiC. A clustering is introduced to obtain compact task-specific causal representations.}
  \label{jiegou}
\end{figure*}

\section{Methodology}~\label{sec3}
\subsection{VAD in Causality Perspective}
As discussed in Sec. \ref{sec1}, we consider that the spatial-temporal features extracted by DeepRL contain both causal variable that determine the normality and label-independent non-causal variable. Existing VAD methods typically use DeepRL to learn the statistical independence of normal videos with redundant and non-helpful representations. In response, we design a structural causal model (SCM) shown in Figure~\ref{motivation}(b) to formulate unsupervised VAD and guide our CRC framework to learn robust knowledge beyond the available training data. The following common cause principle describes the connection between statistical dependence and causality:

\begin{center}
  \fbox{
    \parbox{.95\linewidth}{\S1 \textbf{ Common cause principle:} \textit{If two observables $\bm{X}$ and $\bm{Y}$ are statistically dependent, then there exists a variable $\bm{Z}$ that causally influences both and explains all the dependence in the sense of making them independent when conditioned on $\bm{Z}$.}}}
    \label{s1}
\end{center}

For the unsupervised VAD task, we set $\bm{X}$ and $\bm{Y}$ to denote observable regular events and normality (label 0), respectively. The causal variable $\bm{Z}$, which has a causal effect on both the original data distribution and normality learning, is not directly observable, which may be usual targets (as opposed to unexpected objects in appearance anomalies) or regular object-scene interactions (as opposed to violations in motion anomalies). In this regard, we attempt to use the directed acyclic graph in Figure~\ref{motivation}(b) to implicitly learn a set of causal factors $\{\bm{z}_1, \cdots, \bm{z}_n\}$ with internal consistency for characterizing various types of normal events, as follows:
\begin{align}
  & \bm{X}:=f\left(\bm{Z}, \bm{U}, \bm{P}\right), \bm{Z} \bot  \bm{U} \bot  \bm{P}, \\
  & \bm{Y}:=h\left(\bm{Z}, \bm{P}\right)=h\left(g(\bm{X}), \bm{P}^\prime\right), \bm{P} \bot  \bm{P}^\prime,
\end{align}
where $\bm{U}$ denotes the non-causal variable only affecting $\bm{X}$, e.g., domain-specific information that contributes nothing to normality learning. $\bm{P}$ and $\bm{P}^\prime$ denotes joint-independent unexplained perturbation noise. $f{(\cdot, \cdot, \cdot)}$, $h(\cdot, \cdot)$ and $g(\cdot)$ are regarded as unknown structural functions with causal mechanisms. According to \S1 and the invariant causal mechanism, for any distribution $P(\bm{X}, \bm{Y})\in \mathcal{P}$, when the causal variable $\bm{Z}$ is given, then a general conditional distribution $P(\bm{Y}|\bm{Z})$ must exist. Thus, representations that imply causality are essential for learning normality robust to diverse patterns of normal events. 

However, it is impractical to observe causal variable from unstructured videos, i.e., there is no available prior to guide us to define causal representations dissuasively. Following the consensus in CausalRL, we attempt to encourage the model to learn a set of orthogonal causal factors with the following principle \cite{peters2017elements,scholkopf2021toward}:
\begin{center}
  \fbox{
    \parbox{.95\linewidth}{\S2 \textbf{Independent causal mechanism:} \textit{The causal generative process of a system's variables is composed of autonomous modules that do not inform or influence each other. In the probabilistic case, this means that the conditional distribution of each variable given its causes (i.e., its mechanism) does not inform or influence the other mechanisms.}}}
    \label{s2}
\end{center}
\hyperref[s2]{\S2} inspires us to find the unobservable causal factors $\{\bm{z}_1, \cdots, \bm{z}_n\}$ with the separate intervening nature of $\bm{Z}$ and their independence. Corresponding to the causal factorization of the VAD representations, we know that: (1) $\bm{Z}$ that fully generalize normality are separated from $\bm{U}$, i.e., interventions on  $\bm{U}$ do not change  $\bm{Z}$ and $\bm{Y}$. (2) The causal factors $\{\bm{z}_1, \cdots, \bm{z}_n\}$ is jointly independent, and mechanism $P(\bm{z}_i|PA_i)$ does no inluence or transfer information with $P(\bm{z}_j|PA_j)$ if $j\ne i$, where $PA$ denotes the causal parents. (3) The learned task-specific causal representations are causally sufficient for normality learning to explain all statistical independence between $\bm{X}$ and $\bm{Y}$. Therefore, we can factorize the joint distribution of causal factors into conditional as follows:
\begin{equation}
  P\left(\bm{z}_1, \cdots, \bm{z}_n\right)=\prod_{i=1}^n P\left(\bm{z}_i \mid P A_i\right).
\end{equation}

In this work, we consider both the shared and private features of regular events as a phenotypic form of normality. Due to the diversity of video events, conventional representation learning is challenging to outline the distribution of these multi-view features that point to the same causal variable. It is feasible to learn causal representations implicitly through independence, which is consistent with the following hypothesis \cite{scholkopf2021toward}:
\begin{center}
  \fbox{
    \parbox{.95\linewidth}{\S3 \textbf{Sparse mechanism shift:} \textit{Small distribution changes tend to manifest themselves in a sparse or local way in the causal/disentangled factorization, that is, they should usually not affect all factors simultaneously.}}}
    \label{s3}
\end{center}
which motivates us to learn stable causal variable that can respond sensitively to anomalies through intrinsic consistency while resisting label-independent shifts. For implementation, we construct a prototype decomposer to decompose the original representations into private and shared features and train the causality-inspired characterizer to further represent these features with the same causal factors. Besides, we utilize clustering and similarity constraints to explore the consistency and obtain task-specific representations for unsupervised video anomaly detection.

\subsection{Prototype Learning and Decomposition}
Inspired by the ability of the memory \cite{park2020learning} to record the prototype of normal events and constrain the overgeneralization of the DeepRL, we use memory to construct the prototype learning module and obtain the shared and private features. As shown in Figure~\ref{jiegou}, the memory update process $\mathcal{M}_t \to \mathcal{M}_{t+1}$ illustrate the recording of the general patterns of the spatial-temporal features $\bm{F}\in \mathbb{R}^{H\times W\times C}$. Specifically, the memory pool is a two-dimensional matrix, denoted as $\bm{M}\in \mathbb{R}^{C\times N}$, where $N$ denotes the number of memory entries and determines the information capacity of $\mathcal{M}_t$. The memory pool contains no learnable parameters but updates its memory entries to record normality through the write operation with $\bm{M}$ serving as query $\bm{Q}_M$, as follows:
\begin{equation}
\begin{aligned}
  \mathcal{M}_{t+1} &= l_2\left(\bm{M}+\bm{V}_F \varPsi \left(\frac{\bm{K}_F^T\bm{Q}_M}{\sqrt{C}}\right)\right),\\ \bm{V}_F&=\bm{K}_F=e(\bm{F})\in \mathbb{R}^{C\times \hat{N}},
\end{aligned}
\end{equation}
where $e(\cdot)$ denotes expanding $\bm{F}$ along the spatial dimension so that $\hat{N}=H\times W$. $l_2(\cdot)$ is L2-norm to keep the data scale of $\mathcal{M}_t $ and $ \mathcal{M}_{t+1}$ consistent, and $\varPsi$ denotes softmax. In contrast, the read operation aims to reconstruct $\bm{F}$ as prototype $\bm{F}^\prime$ with expanded $\bm{F}$ serving as query $\mathcal{Q}_F$, as shown in Figure~\ref{jiegou}:

\begin{equation}
  \bm{F}^\prime = \bm{V}_M \varPsi \left(\frac{\bm{K}_M^T e(\bm{F})}{\sqrt{C}}\right), \bm{V}_M=\bm{K}_M=\bm{M}\in \mathbb{R}^{C\times N}.
\end{equation}

Existing work \cite{park2020learning,liu2021hybrid,wang2023memory} assumes that $\bm{F}^\prime$ can effectively detect anomaly differences, i.e., the anomalous event will lose its own patterns and thus encounter significant errors in proxy tasks. However, over-strong memory may make the well-trained model unable to reason about normal events with shifts. Learning the distribution of prototype features is insufficient for describing diverse regular events and discriminating complex anomalies. Therefore, We use CausalRL to further explore the intrinsic connection and understand the inherent differences between positive and negative samples with causal consistency.

The raw features $\bm{F}$ contain shared prototypical semantics and unique personalized semantics, i.e., shared and private features, denoted as $\{\bm{F}_s, \bm{F}_p\}$. As stated in Sec.~\ref{sec1}, both $\bm{F}_s$ and $\bm{F}_p$ are statistically associated with label 1. Referring to sparse representation learning \cite{wu2022self}, we design a SE-like \cite{hu2018squeeze} process to strip $\bm{F}_s$ and $\bm{F}_p$ from $\bm{F}$ and $\bm{F}^\prime$. The details are shown in Figure~\ref{jiegou}. First, $\bm{F}$ and $\bm{F}^\prime$ are average and max pooled, denoted as $\{\bm{f}_\text{avg},\bm{f}^\prime_\text{avg},\bm{f}_\text{max},\bm{f}^\prime_\text{max}\} \in \mathbb{R}^{C}$, which are then mapped to the difference scores $\{\alpha, \beta\}$ by two multi-layer perception (MLP) with learnable parameters $\{\theta_1, \theta_2\}$, as follows:
\begin{equation}
  \alpha = \text{MLP}(\bm{f}_\text{avg}-\bm{f}^\prime_\text{avg}; \theta_1), \beta = \text{MLP}(\bm{f}_\text{max}-\bm{f}^\prime_\text{max}; \theta_2).
\end{equation}
Finally, we use $\alpha$ and $\beta$ to quantitatively filter the prototypical semantics in $\bm{F}$, as follows:
\begin{equation}
  \bm{F}_p = \frac{\alpha+\beta}{2} \circledast \bm{F}, \bm{F}_s = \left(1-\frac{\alpha+\beta}{2}\right) \circledast \bm{F}^\prime,
\end{equation}
where $\circledast$ denotes the channel-wise multiplication.

\subsection{Representation Consistency Learning}
Inspired by principles \hyperref[s1]{\S1} and \hyperref[s2]{\S2}, we know that there exist jointly independent causal factors capable of fully generalizing the statistical dependence from low-level normal videos to high-level normality. Furthermore, \hyperref[s3]{\S3} indicates that the individualized features of normal events have a limited effect on the causal factors and their consistency. Therefore, we construct a causality-inspired characterizer (CiC) to learn unobservable causal variable and model the intrinsic consistency of normal events. Specifically, the spatial-temporal features of $b$ video clips from the same batch are decomposed and fed into CiC, which maps their shared and private features into causal representations: $\bm{R} = \{\bm{r}_1; \bm{r}_2; \cdots; \bm{r}_b\} = \text{CiC}(\bm{F}_s^1, \bm{F}_s^2, \cdots, \bm{F}_s^b)\in \mathbb{R}^{b \times n}$ and $\tilde{\bm{R}} = \{\tilde{\bm{r}}_1; \tilde{\bm{r}}_2; \cdots; \tilde{\bm{r}}_b\} = \text{CiC}(\bm{F}_p^1, \bm{F}_p^2, \cdots, \bm{F}_p^b)\in \mathbb{R}^{b \times n}$. In practice, $n\ll H\times W\times C$. Unsupervised VAD attempts to learn the normality using only regular events so that $\bm{r}_i$ and $\tilde{\bm{r}}_i$ point to the same label. Then, the causal variables should remain causally invariant to the so-called decomposition intervention, i.e., the causal representations of shared and private features should remain close in the causal factor dimension: 
\begin{equation}
  \label{eq8}
  \max \frac{1}{n} \sum_{i=1}^n \frac{\bm{f}_i\tilde{\bm{f}_i}}{\parallel \bm{f}_i\parallel \parallel \tilde{\bm{f}_i}\parallel},
\end{equation}
where $\bm{f}_i$ and $\tilde{\bm{f}_i}$ denote the $i$-th column of $\bm{R}$ and $\tilde{\bm{R}}$, respectively.

By maximizing the similarity of the same $n$ causal factors on shared and private features, we can encourage CiC to learn causal factors that can strip label-independent non-causal variable from redundant deep spatial-temporal features. In addition, to ensure that the causal factors are jointly independent, we construct three correlation matrices on $\bm{R} \to \tilde{\bm{R}}$, $\bm{R}\to \bm{R}$, and $\tilde{\bm{R}} \to \tilde{\bm{R}}$, denoted as $\bm{C}_1$, $\bm{C}_2$, and $\bm{C}_3$, as shown in Figure~\ref{jiegou}. similar to Eq.~\ref{eq8}, the non-diagonal elements of $\bm{C}_1$ are also the cosine similarity between the corresponding columns of $\bm{R}$ and $\tilde{\bm{R}}$. In contrast, $\bm{C}_2$ and $\bm{C}_3$ present the similarity within $\bm{R}$ and $\tilde{\bm{R}}$: $\bm{C}_2(i,j) = \frac{\bm{f}_i{\bm{f}_j}}{\parallel \bm{f}_i\parallel \parallel {\bm{f}_j}\parallel}$ and $\bm{C}_3(i,j) = \frac{\tilde{\bm{f}}_i{\tilde{\bm{f}}_j}}{\parallel \tilde{\bm{f}}_i\parallel \parallel \tilde{\bm{f}}_j\parallel}$. The final optimization objective is to maximize the diagonal elements of the correlation matrix $\bm{C}_1$ (Note that the diagonal elements of $\bm{C}_2$ and $\bm{C}_3$ are constant 1) and minimize the non-diagonal matrices of $\bm{C}_1$, $\bm{C}_2$, and $\bm{C}_3$, as follows:
\begin{equation}
  \label{eq9}
  \min \lambda \parallel \bm{C}_1-\bm{I}\parallel_F^2 + \parallel \bm{C}_2-\bm{I}\parallel_F^2 + \parallel \bm{C}_3-\bm{I}\parallel_F^2,
\end{equation}
where $\lambda$ is a trade-off hyper-parameter, ans $\bm{I}$ denotes the identity matrix. In this way, we can ensure that the causal factors are jointly independent and invariant to the decomposition intervention. According to \hyperref[s3]{\S3}, normal events with shifts are only locally different regarding causal representations. Therefore, we follow \cite{chang2020clustering} to introduce clustering to obtain a tighter causal representation $\bm{R}$ and further enhance the model to discriminate normal events with clustering effects. In addition, we introduce memory separateness and compactness loss \cite{park2020learning} to optimize the memory pool.

\subsection{Anomaly Detection with Causal Consistency}
Due to the special setting of the anomaly detection task, only negative samples are available for training, so the well-trained CRC framework is only effective in decomposing and constructing causal representation consistency for normal events. In the testing phase, we compute anomaly score $s_t$ by measuring the deviations to the learned causal factors in terms of consistency and representations:
\begin{equation}
  \label{eq10}
  s_t = g(\parallel \bm{C}_1-\bm{I} \parallel_F^2 \times D),
\end{equation}
where $g(\cdot)$ denotes the max-min normalization over all frames. $D$ is the clustering distance between the causal representations of the input video clip and the cluster center. The former part of $s_t$, $\bm{C}_1-\bm{I}$, discriminates anomalies by the consistency in the causal variable, while the latter $D$ measures the distance to the normal representation. 

\begin{figure*}
  \centering
  \includegraphics[width=\textwidth]{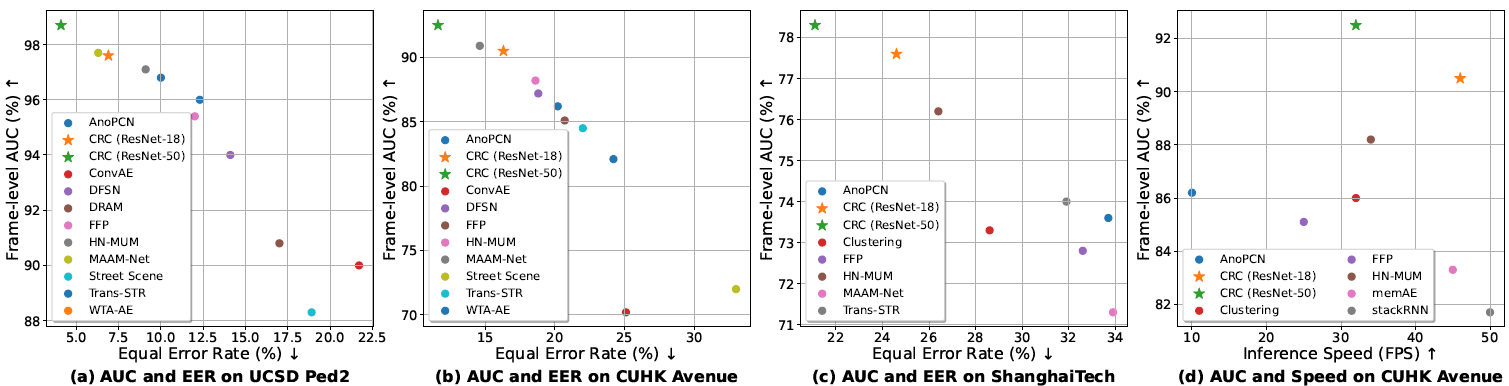}
  \caption{Quantitative performance comparison. (a)-(b) show the frame-level AUC and EER of our method (marked by pentagrams) with existing methods (marked by circles) on the three datasets, respectively, while (d) presents the inference speed on the CUHK Avenue dataset. $\uparrow$ denotes that larger values indicate better performance, while $\downarrow$ vice versa. Best viewed in color.}
  \label{scatter}
\end{figure*}

\section{Experiments}
\subsection{Experimental Setup}
\subsubsection{Datasets}
We conduct extensive experiments to validate the effectiveness of the proposed CRC framework on three leading unsupervised VAD benchmarks, including UCSD Ped2 \cite{ped2}, CUHK Avenue \cite{avenue}, and ShanghaiTech \cite{shanghai}. All training sets are normal videos collected from the real world, while anomalous events from similar scenes are only available to the test set. \textit{UCSD Ped2} \cite{ped2} is a small-scale dataset containing 16 training and 12 test videos, captured from the university campus. As an early unsupervised VAD benchmark, its scenario is simple, with regular samples walking normally on the sidewalk, while the anomalous events include riding bikes, skateboarding, and driving. \textit{CUHK Avenue} \cite{avenue} is a large-scale single-scene VAD dataset. The training and test sets contain 21 and 16 videos with 47 anomalous events. The collectors simulated appearance-only (e.g., the person on the lawn), motion-only (e.g., loitering and wandering), and appearance-motion anomalies (e.g., papers being scattered), making CUHK Avenue more challenging. \textit{ShanghaiTech} \cite{shanghai} is the most challenging benchmark, collecting 130 anomalies from 13 scenes. The data size and cross-scene nature make it difficult for unsupervised methods to learn effective deep representations to describe diverse normal events.

\subsubsection{Evaluation Metrics}
In the testing phase, we measure the input samples against the learned causality-inspired normality to calculate the degree of abnormality and output a continuous anomaly score in the range $[0,1]$. A high score indicates that the more likely the test sequence is to be anomalous. In contrast, the given labels are binary discrete, where $0$ indicates normal and $1$ indicates abnormal. Following previous work \cite{liu2018future,bao2022hierarchical}, we calculate the true-positive-rate and false-positive-rate at multiple thresholds and plot the receiver operating characteristic curve, using the area under the curve (AUC) as the primary evaluation metric to present the effectiveness of our method for anomaly detection. In addition, the equal error rate (EER) is used as a complementary metric to demonstrate the robustness of the CRC framework, which is compared with available methods. With the same implementation platform, we report the average inference speed of our method on the CUHK Avenue \cite{ped2} dataset to validate its deployment potential on resource-limited terminal devices.

\subsubsection{Implementation Details}
We use the PyTorch \cite{paszke2019pytorch} framework to implement the proposed method on an Nvidia 3090 GPU. The Adam \cite{kingma2014adam} optimizer is used to train the model with an initial learning rate of $8\times 10^{-5}$. The batch size $b$ is set to 8. In the initial stage, we remove the clustering constraints and optimize the characterizer without clustering. After 100 epochs, we compute the clustering centers using K-means and update them with an alternating optimization \cite{chang2020clustering}. The video frames are resized to $224\times 224$ pixels. The feature extractor is a 5-layer convolutional encoder. The two MLPs in the prototype decomposer are three-layer fully connected neural networks with sigmoid activation in the output layer. We select ResNet-18 and ResNet-50 \cite{he2016deep} as the backbone of CiC. The trade-off hyper-parameter $\lambda$ in Eq. \ref{eq9} is set to 10, 18, and 20 for UCSD Ped2 \cite{ped2}, CUHK Avenue \cite{avenue}, and ShanghaiTech \cite{shanghai}, respectively.

\subsection{Comparisons with State-of-the-art Methods}

\begin{table}[]
  \centering
  \caption{Results of the frame-level AUC comparison.}
  \label{t1}
   \resizebox{.47\textwidth}{!}{
  \begin{threeparttable} 
  \begin{tabular}{@{}clccc@{}}
  \toprule
  \multirow{2}{*}{\textbf{Type}}   & \multicolumn{1}{c}{\multirow{2}{*}{\textbf{Method}}} & \multicolumn{3}{c}{\textbf{Frame-level AUC (\%)}} \\ \cmidrule(l){3-5} 
 & \multicolumn{1}{c}{}  & UCSD Ped2& CUHK Avenue& ShanghaiTech    \\ \midrule
  \multirow{6}{*}{\rotatebox{90}{Traditional}}& MPPCA \cite{kim2009observe} & 69.3   & -    &  -   \\
 & MPPC+SFA \cite{kim2009observe} & 61.3   &  -   & -    \\
 & MDT \cite{ped2}  & 82.9   &  -   &  -   \\
 & AMDN \cite{xu2017detecting} & 90.8   &   -  &  -   \\
 & Unmasking \cite{tudor2017unmasking} & 82.2   & 80.6&  -   \\
 & MT-FRCN \cite{hinami2017joint}  & 92.2   &  -   & -    \\
 \midrule
  \multirow{18}{*}{\rotatebox{90}{Deep Learning-based}} 
 & ConvAE \cite{hasan2016learning}   & 90.0 & 70.2&  -   \\
 & ConvLSTM-AE \cite{lu2019future} & 88.1   & 77.0  &  -   \\
 & AMC \cite{nguyen2019anomaly} & 96.2   & 86.9&  -   \\
 & FFP \cite{liu2018future}  & 95.4   & 85.1& 72.8\\
 & MemAE \cite{gong2019memorizing} & 94.1   & 83.3& 71.2\\
 & AnoPCN \cite{ye2019anopcn}   & 96.8   & 86.2& 73.6\\
 & Mem-Guided \cite{park2020learning} & 97.0 & 88.5& 70.5\\
 & AMMC-Net \cite{cai2021appearance}  & 96.6   & 86.6& 73.7\\
 & Clustering \cite{chang2020clustering} & 96.5   & 86.0 & 73.3\\
 & TAC-Net \cite{huang2021abnormal} & 98.1   & 88.8& 76.5\\
 & STD \cite{chang2022video} & 96.7   & 87.1& 73.7\\
 & STC-Net \cite{zhao2022exploiting}  & 96.7   & 87.8& 73.1\\
 & STM-AE \cite{liu2022learning}  & 98.1   & 89.8 & 73.8 \\
 & Bi-Prediction \cite{chen2022comprehensive}   & 97.4   & 86.7& 73.6\\
 & HSNBM \cite{bao2022hierarchical} & 95.2   & 91.6& 76.5\\ 
 & MAAM-Net \cite{wang2023memory} & 97.7   & 90.9& 71.3\\ \cmidrule(l){2-5} 
 & CRC (ResNet-18)   & 97.6   & 90.5& 77.6\\
 & CRC (ResNet-50)    & \textbf{98.7}   & \textbf{92.5}& \textbf{78.3}\\ \bottomrule
  \end{tabular}
\begin{tablenotes} 
    \item \small Bold numbers indicate the best performance.
\end{tablenotes} 
\end{threeparttable} 
   }
\end{table}

We perform quantitative comparisons with traditional handicraft feature-based \cite{kim2009observe,ped2} and DeepRL-based methods \cite{liu2018future,gong2019memorizing,liu2022learning,ye2019anopcn,bao2022hierarchical} to demonstrate the effectiveness of the proposed CRC framework. The results are shown in Table~\ref{t1} and Figure~\ref{scatter}. Among them, Table~\ref{t1} presents the frame-level AUCs of existing unsupervised VAD methods on three mainstream datasets, and our CRC framework achieves 98.7\%, 92.5\%, and 78.3\% AUCs on UCSD Ped2 \cite{ped2}, CUHK Avenue \cite{avenue} and ShanghaiTech \cite{shanghai} datasets, respectively, outperforming other methods. Compared with earlier manual feature-based methods, deep learning methods achieve significant performance gains due to the powerful representational learning capability of deep neural networks. However, they fail to process the complex cross-scene ShanghaiTech \cite{shanghai} dataset, and the performance is limited under the unsupervised setting. The proposed CRC framework pioneers the causal representations into unsupervised normality learning, which attempts to mitigate the negative impact of diverse normal events from a causal perspective and discriminates anomalies with consistency, achieving an AUC gain of 1.8\% on the ShanghaiTech dataset. For complex real-world videos, the learned normality with intrinsic causality by CausalRL is more effective than conventional representation learning. 

In addition, we show the EERs and inference speed of the proposed method in Figure~\ref{scatter}, where (a)-(c) show the AUC and EER results on three publicly available benchmarks, while (d) visually compares the performance and inference speed on the CUHK Avenue \cite{ped2} dataset. In addition to those already cited in Table~\ref{t1}, other methods involved in the comparison include DRAM \cite{xu2015learning}, WTA-AE \cite{tran2017anomaly}, stackRNN \cite{shanghai}, DFSN \cite{ramachandra2020learning}, Street Scene \cite{ramachandra2020street}, Trans-STR \cite{sun2023transformer}, and HN-MUM \cite{li2023hn}. Our CRC framework implemented using ResNet-50 achieves EERs of 4.1\%, 11.6\%, and 21.1\% on the UCSD Ped2 \cite{ped2}, CUHK Avenue \cite{avenue}, and ShanghaiTech \cite{shanghai} datasets. The average inference speed of CRC (ResNet-18) and CRC (ResNet-50) is around 46 FPS and 32 FPS, respectively, which means that they take 0.022s and 0.031s from video read-in to anomaly score output, meeting the demand of real-time detection. Although the inference speed of stackRNN \cite{shanghai} is faster than the proposed method, our CRC (ResNet-18) and CRC (ResNet-50) show an advantage in detection accuracy, with AUC improvements of 8.8\% and 10.8\% (90.5\% \& 92.5\% vs. 81.7\%), respectively.

\subsection{Ablation Study}

\begin{table}[]
  \centering
  \caption{Results of ablation study.}
  \label{t2}
\resizebox{.47\textwidth}{!}{
  \begin{threeparttable}
\begin{tabular}{cccccccccc}
\toprule
\multirow{2}{*}{ID} & \multicolumn{3}{c}{\textbf{Component}}& \multicolumn{3}{c}{\textbf{Constraint}}    & \multicolumn{3}{c}{\textbf{Frame-level AUC (\%)}} \\ 
\cmidrule(l){2-4} \cmidrule(l){5-7} \cmidrule(l){8-10}
& $\mathcal{C}$ & AP & MP& $ \bm{C}_{1(F)}$ & $\bm{C}_{2(F)}$ & $\bm{C}_{3(F)}$ & Ped2 & Avenue & S.T.\\ \midrule
1& \XSolidBrush   & \CheckmarkBold & \CheckmarkBold & \CheckmarkBold& \CheckmarkBold& \CheckmarkBold& 91.6 & 83.2 & 72.4 \\
2& \CheckmarkBold & \XSolidBrush   & \CheckmarkBold & \CheckmarkBold& \CheckmarkBold& \CheckmarkBold& 96.9 & 89.6 & 77.1 \\ 
3& \CheckmarkBold & \CheckmarkBold & \XSolidBrush   & \CheckmarkBold& \CheckmarkBold& \CheckmarkBold& 97.1 & 89.4 & 76.6 \\
4& \CheckmarkBold & \CheckmarkBold & \CheckmarkBold & \XSolidBrush& \CheckmarkBold& \CheckmarkBold& 89.1 & 81.9 & 70.7 \\
5& \CheckmarkBold & \CheckmarkBold & \CheckmarkBold & \CheckmarkBold& \XSolidBrush& \CheckmarkBold& 96.3 & 88.2 & 76.1 \\
6& \CheckmarkBold & \CheckmarkBold & \CheckmarkBold & \CheckmarkBold& \CheckmarkBold& \XSolidBrush& 96.6 & 89.1 & 75.8 \\
7& \CheckmarkBold & \CheckmarkBold & \CheckmarkBold & \CheckmarkBold& \XSolidBrush& \XSolidBrush& 96.1 & 88.1 & 75.3 \\
8& \CheckmarkBold & \CheckmarkBold & \CheckmarkBold & \CheckmarkBold& \CheckmarkBold& \CheckmarkBold& \textbf{97.6} & \textbf{90.5} & \textbf{77.6} \\ \bottomrule
\end{tabular}
\begin{tablenotes} 
  \item \small $\mathcal{C}$: Clustering; AP/MP: Average/Max Pooling; $\bm{C}_{i(F)}$=$\parallel \bm{C}_i-\bm{I}\parallel_F^2$, $i=\{1,2,3\}$.
\end{tablenotes} 
\end{threeparttable} 
}

\end{table}

To validate the impact of individual components and optimization constraints on causality-inspired normality learning and their effectiveness in the VAD task, we conduct an ablation study and quantitatively compared the frame-level AUC of each model variant, as shown in Table~\ref{t2}. Model 1, which removes the clustering module and learns the causal factors by characterizing consistency only, suffers significant performance degradation on all three benchmarks. Unlike the classification task, VAD has only negative samples available during the training phase and cannot construct classifiers to encourage the model to learn task-specific representations, so we introduce clustering to characterize normal events as intrinsic to causal representations. The causal characterizer is optimized to learn task-specific representations valid for video anomaly detection by updating the clustering centers, yielding remarkable improvement for the UCSD Ped2 \cite{ped2}, CUHK Avenue \cite{avenue}, and ShanghaiTech \cite{shanghai} datasets with AUC gains of 6.0\%, 7.3\% and 5.2\%. 

Models 2 \& 3 compare the contribution of average pooling and maximum pooling in the prototype decomposer. The performance gap indicates that average pooling aggregates global information and separates shared and private features more effectively. Furthermore, both pooling strategies contribute to normality learning with cumulative gains when compared to the full framework in Model 8. As stated in Sec.~\ref{sec3}, we are inspired by \hyperref[s1]{\S1} and \hyperref[s2]{\S2} to use constraints on the correlation matrix to encourage the model to learn a set of independent causal factors. The impact of each constraint on performance is shown in Models 4-7. Model 4 suffers the overall and most severe AUC decline, indicating that the $\bm{R}\to \tilde{\bm{R}}$ correlation constraint, i.e., $\bm{C}_1$, is critical for causal representation learning. In contrast, the other two constraints of the representation matrix, i.e., $\bm{C}_2$ in model 5 and $\bm{C}_3$ in model 6, bring only minor gains for video anomaly detection. Even ignoring the corresponding constraints under limited computational resources, the impact on model performance is limited, as shown in Model 7.

\subsection{Sensitivity Analysis}
\begin{figure}
  \centering
  \includegraphics[width=.49\textwidth]{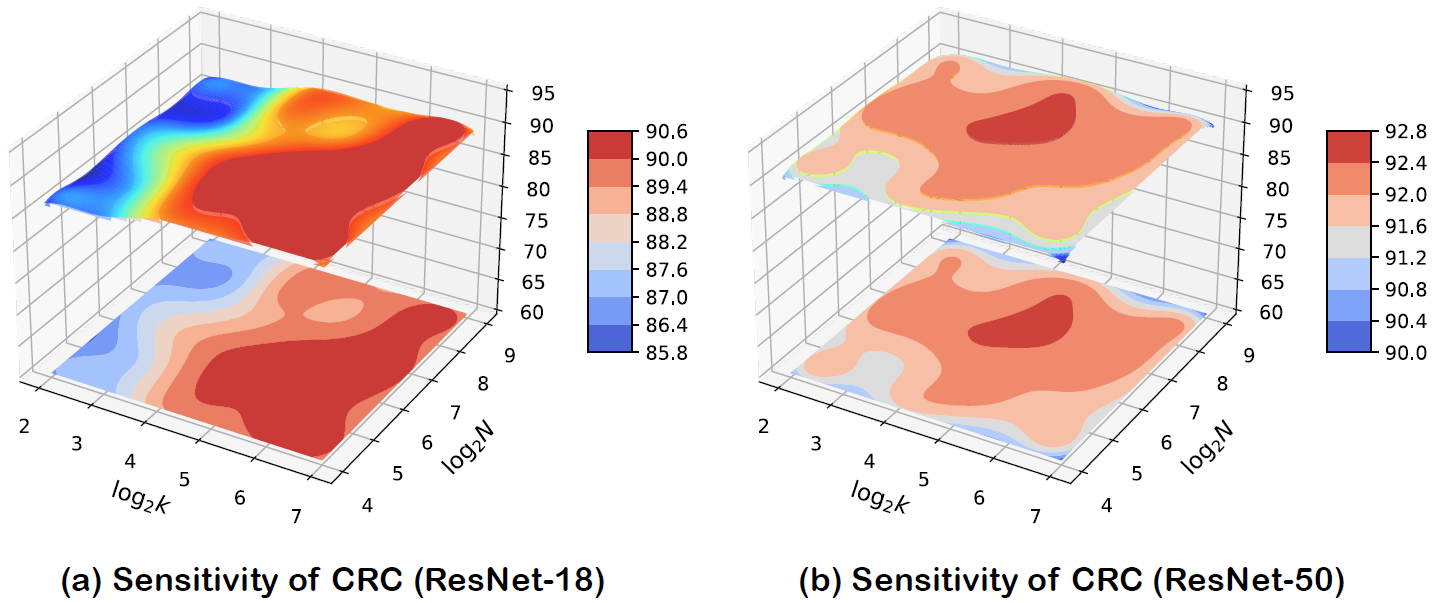}
  \caption{Results of sensitivity analysis on $N$ and $k$. For better visual effects, we interpolated the original data during the mapping. Best viewed in color.
  }
  \label{sensitivity}
\end{figure}

\begin{figure}[t]
  \centering
  \includegraphics[width=.48\textwidth]{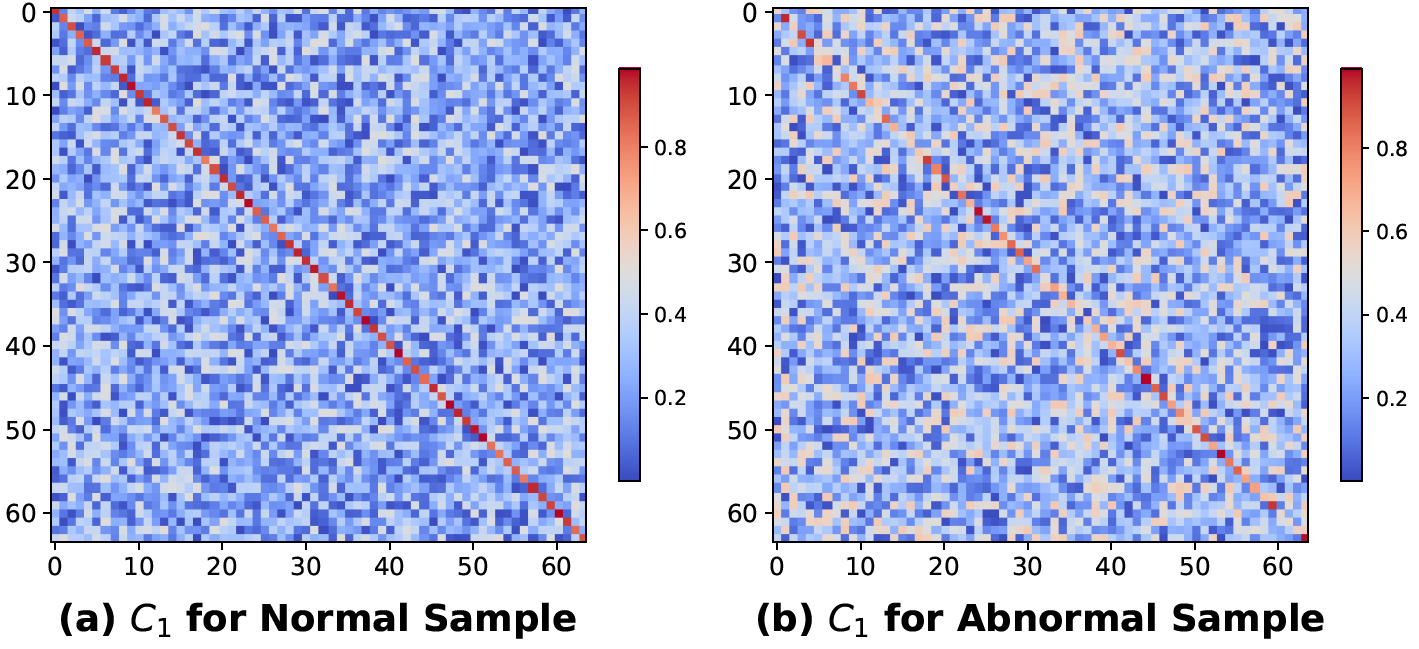}
  \caption{$\bm{R}\to \tilde{\bm{R}}$ correlation matrix visualization.}
  \label{correlation}
\end{figure}

We conduct extension experiments on CUHK Avenue \cite{avenue} to explore the sensitivity of model performance to the number of memory entries and clustering centers, as shown in Figure~\ref{sensitivity}, where (a) is implemented with ResNet-18 while (b) is with ResNet-50. Specifically, we test the AUC performance under different parameter settings: $N=\{16, 32, 64, 128, 256, 512\}$ and $k=\{4, 8, 16, 32, 64, 128\}$. As stated in Sec.~\ref{sec3}, $N$ determines the information capacity of the memory pool. A large $N$ may make the recorded memory entries untypical, while a too-small $N$ will cause the memory pool to lose prototypical features. The results in Figure (4) show that choosing an appropriate $N$ is necessary for prototype decomposition. The clustering is used to make representations compact, and $k$ is critical for representation learning and downstream anomaly detection. Figure 4(b) clearly demonstrates that the impact of $k$ is twofold: too small a $k$ may make it difficult for the CRC framework to find the appropriate cluster centers that make the causal representation tight, while too large a $k$ may make learned cluster centers close to the anomalous events. The effects $N$ and $k$ on the model performance are joint. Our CRC framework achieves the best performance at $\{N=64, k=128\}$ and $\{N=32, k=128\}$ on the CUHK Avenue \cite{avenue} with ResNet-18 and ResNet-50. In addition, the performance shown in Figure~\ref{sensitivity}(b) is more stable than that in \ref{sensitivity}(a), indicating the superior robustness of ResNet-50 compared to ResNet-18.

\subsection{Feasibility Analysis}

The proposed CRC framework attempt to address unsupervised VAD via causal representation learning and detect anomalies with the learned consistency of the causal variables for regular events. To qualitatively present the response of causal representation consistency to anomalies, we randomly select a normal and abnormal sample from the test set of ShanghaiTech \cite{shanghai} dataset and partially visualized their $\bm{R}\to \tilde{\bm{R}}$ correlation matrices, as shown in Figure~\ref{correlation}. The causal factors of the regular events in (a) show good independence, i.e., the diagonal elements of the matrix are close to 1 while the other elements are as small as possible. In contrast, many non-diagonal elements of the matrix for abnormal events in (b) are greater than 0.4, indicating that the causal factors learned on negative samples fail to represent the abnormal patterns. Therefore, we can quantitatively measure the deviation of the test samples from the learned representation consistency by calculating the Frobenius norm (F-norm) of the correlation matrix with the identity matrix, i.e., $\parallel \bm{C}_1-\bm{I} \parallel_F^2$. The F-norm of the two selected samples are 18.3 and 22.4, respectively, which can amplify the score gap between regular and abnormal events, as presented in Eq.~\ref{eq10}.

\subsection{Temporal Localization}
\begin{figure}[t]
  \centering
  \includegraphics[width=.48\textwidth]{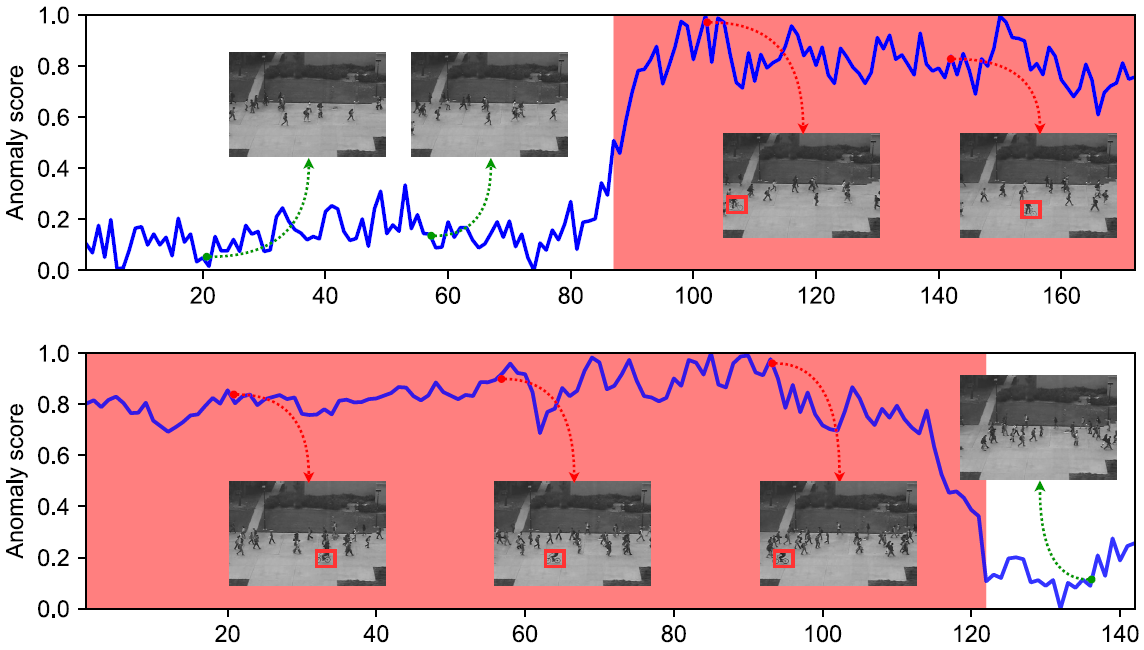}
  \caption{Results of temporal localization for anomalies.}
  \label{curve}
\end{figure}

Moreover, we plot the score curves of two sample videos from the UCSD Ped2 \cite{ped2} dataset to qualitatively verify the quick response of our method to abnormal events, as shown in Figure~\ref{curve}. For the regular interval, the anomaly scores fluctuate slightly but always remain low (generally $<$ 0.2). When anomalous events occur, the anomaly score rises rapidly and remains high (generally $>$0.6) until the anomaly ends or leaves the field of view, indicating that our CRC framework can quickly respond to anomalies and provide accurate temporal localization for abnormal events.

\section{Conclusion}
In this paper, we address unsupervised video anomaly detection from a causal perspective and propose causally-inspired representation consistency learning to model video normality with intrinsic causality. The proposed CRC framework exploits causal principles to mine unobservable causal factors that can fully characterize normality and discriminates anomalous events with the learned representation consistency. Extensive experimental results on three benchmarks validate the effectiveness and superiority of causal representation learning on video anomaly detection. The learned normality-specific causal variable with inherent consistency can effectively reason about regular events with label-independent bias and respond quickly and sensitively to real-world anomalies. In future work, we will further explore potential causal mechanisms in unsupervised normality learning and develop robust video anomaly detection models that can bridge the domain gaps in multi-scene and multi-view real-world videos.

\section*{Acknowledgements}
This work is funded by the China Mobile Research Fund of MoE (Grant No. KEH2310029), NSFC (Grant No. 62250410368), and the Specific Research Fund of the Innovation Platform for Academicians of Hainan Province (Grant No. YSPTZX202314). This work is also supported by the Shanghai Key Research Lab of NSAI and the Joint Lab on Networked AI Edge Computing Fudan University-Changan.

\balance

\bibliographystyle{IEEEtran}
\bibliography{refs.bib}

\end{document}